 \documentstyle[12pt]{article}
 
\begin{document}

\title{Transformations and BRST-Charges in $2+1$ Dimensional Gravitation}
\author{G\'eza F\"ul\"op\\ Institute of Theoretical Physics\\
S-412 96 G\"oteborg\\ Sweden}
\date{September 1992}

\maketitle

\vskip -10cm
\noindent \hskip 10cm G\"oteborg,ITP 92-39
\vskip 2mm
\noindent \hskip 10cm   September,1992
\vskip +10cm

\begin{abstract}

   Canonical transformations relating the variables of the ADM-, Ash\-te\-kar's
and Witten's formulations of gravity are computed in
$2+1$~dimensions. Three different forms of the BRST-charge are given
in the $2+1$ dimensional Ashtekar formalism, two of them using Ashtekar's
form of the constraints and one of them using the forms suggested by
Witten. The BRST-charges are of different rank.
\end{abstract}

\section{Introduction}

   There exist two different formalisms describing gravitation
in $3+1$ dimensions: the ADM- and Ashtekar's formalisms. There is a canonical
transformation relating these to each other as shown by A.Ashtekar
\cite{ashtekar1}, J.Friedman and I.Jack \cite{friedmann} and M.Henneaux,
J.E.Nelson and C.Schomblond \cite{henneaux}. This transformation is
built on the properties of the spatial spin connection.

   In $2+1$ dimensions there are already three different formulations:
the ADM, Ashtekar's and Witten's \cite{Witten}. These formulations are
 equivalent if the metric is nondegenerate.
For the relations between them see e.g. \cite{mano},
\cite{Ingemar's article}. One of the purposes
of this paper is to find an explicit canonical transformation between
the Ashtekar and the ADM formalisms in $2+1$ dimensions. In this case
we can not make use of the spatial spin connection, because the fundamental
variables in the Ashtekar's formalism have SO(1,2) indices while
the spatial spin connection has SO(2) indices.(Section 2).

   Since the constraint algebra in Ashtekar's formulation is different
from the one in Witten's formulation a canonical transformation relating
these formulations must take place in the extended phase space
and  involve ghost fields too \cite{fradkin}.
Moreover the constraint algebra of the Ashtekar formalism shows some
peculiar features: the Poisson bracket between the Hamiltonian and
the vector constraints contains terms of the form:  ${\cal H}{\cal G}_I$
and  ${\cal H}_a{\cal G}_I$ , where  ${\cal H}$ is the Hamiltonian constraint,
${\cal H}_a$ ($a=1,2$) are the vector constraints and the ${\cal G}_I$'s
are the constraints called Gauss' law. These terms can be interpreted
in two different ways: one can either say that ${\cal H}{\cal G}_I$
is the Hamiltonian constraint multiplied by a structure function
which is `accidentally' a function of Gauss' law, or that it is
Gauss' law multiplied by the structure function containing the Hamiltonian
constraint. These two interpretations give birth to two different
forms of the BRST-charge of the theory. Together with the BRST-charge of the
Witten's formalism one has three different forms of the BRST-charge of the
same theory. It is possible to obtain one form knowing  another one
by canonical transformations. The purpose of Section 3 is to show
these forms of the BRST-charge explicitly and to compute one of the
canonical transformations between them.

   It is worth noting that these forms of the BRST-charge while belonging
to the same theory are of different rank.

\section{The transition between\-
the ADM-\- and \- Ashtekar's formalisms\-
 in $2+1$ dimensions}

   In the ADM formalism the fundamental variables one uses are the
spatial metric $g_{ab} (a,b = 1...d-1)$ and it's canonically conjugated
momenta $\Pi^{ab}$. An equivalent description of gravity can be given
using as basic variables the triads $e_{aI}$, where `$a$' are the spatial
 indices and `$I$' the internal indices. The internal algebra is SO(3) in
 $3+1$~dimensions and SO(1,2) in $2+1$~dimensions. The relation between
 these variables and the metric is given by:

\begin{equation}
g_{ab} = e_{aI} e_b^I
\end{equation}

The indices `I' are raised and  lowered by the Killing metric of the algebra.

   The canonical momenta conjugated to the triads are  $\Pi^a_I$.
 One can build the same gravitational theory on these variables as
the one built on the $g_{ab}$
and  $\Pi^{ab}$, if an extra constraint is introduced.

   One can make a canonical transformation starting from the triad variables
and obtain a new formalism. See e.g. \cite{Ingemar's book}.

\[ (e_a^I, \Pi^b_J) \rightarrow (K_a^I, E^b_J) \]
where: $ E^b_I = e e^b_I $, $ e = \sqrt{g} $, $g$ being the determinant of
the spatial metric and $ K_a^I $ is related to the extrinsic curvature by:

\[ K_{ab} = e_{aI} K_b^I \]
The extrinsic curvature is weakly symmetrical.

   Ashtekar's formalism uses the fundamental canonical variables $A_a^I$ and
$E^b_I$, where  $A_a^I$ is related to the space component of the spacetime
spin connection:

\[  A_a^I = f^I_{~JK} {\omega}^{JK}_a \]
There exists a canonical transformation relating the formalism using the
variables $K_a^I$ and $E^b_I$ to Ashtekar's formalism.

   The Hamiltonian constraint pushes the spacelike hypersurface forward in
time and the change of the spatial metric of the hypersurface is given
by the extrinsic curvature:

\begin{equation}
\{g_{ab},{\cal H}[M] \} = 2 M K_{ab}
\end{equation}
where  \( {\cal H}[M] \equiv \int M {\cal H} \). If one introduces the
notation:
\[ q^{ab} = E^a_I E^{bI} \]
eq. (2)  gives:
\begin{equation}
\{ q^{ab}, {\cal H}[N] \} = -2 M g^{-1} ( q^{ac}  q^{bd} K_{cd} -
 q^{ab}  q^{cd} K_{cd} )
\end{equation}
with $ N = M /\sqrt{g}$. On the other hand one knows that in Ashtekar's
 formalism the Hamiltonian constraint looks like:

 \begin{equation}
{\cal H}= \frac12 f^{IJ}_{~~K} E^a_I E^b_J F^K_{ab}
\end{equation}
where
$ F^K_{ab} = {\partial}_a A_b^K -{\partial}_b A_a^K + f^K_{~LM} A_a^L A_b^M $
is the curvature field strength. One can use this form of the Hamiltonian
constraint to compute explicitly:
\begin{equation}
\{ q^{cd}, {\cal H}[N] \} = -2 N f^{IJ}_{~~L} E^a_I E^{(c}_J D_a E^{d)L}
\end{equation}
where $ D_a$ denotes the covariant derivative on the phase space:

\[ D_a E^b_I = {\partial}_a E^b_I + f_{IKM} A^K_a E^{bM} \]
and "(~~~)" denotes symmetrization with factor $1/2$ included.

Comparing these two relations one can get the transformation rule between
 $K^M_a$ and $A^M_a$ :

\begin{equation}
K^{M}_{a} = f^{M~I}_{~L} E^b_I q_{ad} D_b E^{dL}
 - f^{IJ}_{~~L} E^b_I E^M_a E_{fJ} D_b E^{fL}    \label{kaold}
\end{equation}
where $ E^M_a $ is defined as an `inverse' to $E^a_M$ :
$E^M_a E^b_M = {\delta}_a^b$.
The `momenta' $E^a_I$ are of course the same for both formalisms:

\begin{equation}
E^b_I = E^b_I  \label{ee}
\end{equation}

If one wants to check whether this transformation expressed by
 equation~\ref{kaold} and \ref{ee} is a canonical one it
is straight forward to obtain:

\[ \{K_a^J(x), E^b_I(x')\} = {\delta}_a^b {\delta}_I^J \delta(x-x') \]

To get however $ \{K_a^J(x), K_b^I(x')\} $ is a very lengthy calculation.

 In $3+1$ dimensions one can obtain a much simpler relation between $A_a^I$
and $K_a^I$ using the spatial spin connection ${\Gamma}_a^I$ which is defined
to satisfy:

\[ {\nabla}_a e^b_I - {\Gamma}^J_{~Ia} e^b_J = 0 \]
with \(  {\Gamma}_a^I = f^{IJK}  {\Gamma}_{JKa} \). The relation between
  $A_a^I$ and $K_a^I$ becomes:

\begin{equation}
 A_a^I = i K_a^I + {\Gamma}_a^I \label{kanew}
\end{equation}
See for example \cite{ashtekar1},\cite{friedmann},\cite{henneaux}.
The relation between the "momenta" of these formulations is of course the same
as given in equation~\ref{ee}. The transformation given by
equations~\ref{kanew}and \ref{ee} is a canonical one since the spatial spin
connection can be
expressed as the functional derivative of a functional depending on the
 "momenta":

\[ {\Gamma}^I_b = \frac{\delta F(E)}{\delta E^b_I} \]
where:
\[  F(E) = 2i \int dx E^a_I {\Gamma}^I_a \]
and the transformations of the form: \( A \rightarrow A + {\partial}_E f(E),
E \rightarrow E \) are canonical.

In $2+1$ dimensions one can not obtain a similar relation to \ref{kanew} and
\ref{ee}: the spatial spin connection is not useful in this case because
it has SO(2) indices, while the fundamental variables carry SO(1,2) indices.

\section{The BRST-Charge}

There exist two different sets of constraints in $2+1$ dimensional
 gravity which give the same description if the spatial metric is
 nondegenerate \cite{mano}, \cite{Ingemar's article}.

   The first set of constraints obey the constraint algebra of general
 relativity and it is determined by geometrical considerations. See for
 example \cite{HKT}.
 In 2+1 dimensions this happens in the same way as in 3+1 dimensions.
 The spatial translations are generated by the vector-constraints. Using
 Ashtekar's formulation these constraints can be expressed as
 \begin{equation}
 {\cal H}_a = E^b_I F^I_{ab} \approx 0
 \end{equation}

 The timelike translations are generated by the Hamiltonian constraint:
 \begin{equation}
{\cal H} = \frac12 f^{IJ}_{~~K} E^a_I E^b_J F^K_{ab} \approx 0
\end{equation}
 We also have the first class constraint called Gauss' law:
 \begin{equation}
 {\cal G}_I \equiv D_a E^a_I = ( \partial_a \delta^J_I +
f_{IK}^{~~J} A^K_a) E^a_J \approx 0
 \end{equation}
 The algebra satisfied by these constraints looks like:
\begin{equation}
\{ {\cal H}_a [N^a], {\cal H}_b [M^b] \} = {\cal H}_a [\pounds_N M^a]  +
 {\cal G}_I [M^a N^b  F^I_{ab} ]
 \end{equation}
 \begin{equation}
 \{ {\cal H}_a [N^a],{\cal H} [M] \} = {\cal H} [ \pounds_N M] +{\cal G}_I
 [M N^a f^{IJ}_{~~K}  E^b_I F^K_{ab}]
 \end{equation}
 \begin{equation}
\{ {\cal H} [N], {\cal H} [M] \} =
 \pm {\cal H}_a [(N \partial_b M - M \partial_b N) q^{ba} ]
 \end{equation}
 \begin{equation}
\{ {\cal G}_I [N^I],{\cal G}_J [N^J] \} = {\cal G}_I [f^{I}_{~JK} N^J N^K]
 \end{equation}
where $ {\cal H}_a [N^a] = \int  {\cal H}_a N^a $, etc.
   $ q^{ba}$ is the spatial metric
    and the sign in the third equation is $-$ for a space-time with Euclidean
signature and it is $+$ for Lorentzian signature.

The other set of constraints wich can  be used in $2+1$ dimensions was
suggested by Witten \cite{Witten}
\begin{equation}
{\cal G}_I =D_a E^a_I  \approx 0
 \end{equation}

\begin{equation}
{\Psi}^I ={\epsilon}^{ab}  F^I_{ab}  \approx 0
 \end{equation}
 The algebra satisfied by these constraints:
\begin{equation}
\{ {\cal G}_I [N^I], {\cal G}_J [M^J] \} = {\cal G}_I [f^I_{~JK} N^J M^K]
 \end{equation}
\begin{equation}
\{ {\Psi}^I [N_I],{\cal G}_J [M^J] \} =
{\Psi}^I [f_{IJK} N^J M^K]
\end{equation}
\begin{equation}
\{{\Psi}^I [N_I],{\Psi}^J [M_J] \} =0
\end{equation}

As it is described in \cite{Ingemar's article}
 these two sets of constraints are related by:
\begin{equation}
{\cal H}_a = \frac {{\epsilon}_{ab}}2  E^b_I {\Psi}^I
\end{equation}
\begin{equation}
{\cal H} = \frac14 f_I^{~JK} E^a_J  E^b_K {\Psi}^I
\end{equation}
or using a symbolical notation:
\begin{equation}
{\cal H}_{\mu} = M_{\mu I} {\Psi}^I
\end{equation}
where $ \mu= 0, 1, 2 $ and $ {\cal H}_{0} \equiv {\cal H} $. This notation  is
a very comfortable one but one shouldn't confuse it with the usual 3-vector
notation, because the quantity $ \{ {\cal H}, {\cal H}_a \} $ is not a
3-vector.So $\mu$ is not a vector index.
We have then the `matrix' $M$ with:
\begin{equation}
M_{a I}= \frac{{\epsilon}_{ab}}2  E^b_I
\end{equation}
\begin{equation}
M_{0 I}= \frac{{\epsilon}_{ab}}4 f_I{}^{JK} E^a_J  E^b_K {\Psi}^I =
\frac12 f_I^{~JK}  E^a_J M_{a K}
\end{equation}
   This matrix is invertible whenever det $ q^{ab}\neq 0 $, and one can write
\[ {\Psi}^I = M^{aI} {\cal H}_a + M^{0I} {\cal H} \]
 Actually det $ q^{ab}= 8  $ det $M$. Using these relations one can rewrite the
 constraint-algebra of the first
(Ashtekar's) set of constraints in the following way:

\begin{equation}
\{ {\cal H}_a [N^a],{\cal H}_b [P^b] \}=
{\cal H}_a [{\pounds}_N P^a] + {\cal G}_I {\cal H}_{\mu} [ (M^{-1})^{\mu I}
 \frac{{\epsilon}_{ab}}2 P^a N^b] \label{alg1}
\end{equation}
\begin{equation}
\{ {\cal H}_a [N^a], {\cal H} [P] \} = {\cal H} [\pounds_N P]+
{\cal H}_{\mu} {\cal G}_I [(M^{-1})^{\mu K} P N^a f^{IJ}_{~~K} E^b_J
 \frac{{\epsilon}_{ab}}2]      \label{alg2}
\end{equation}
The other equations remain the same. One can notice a very interesting
feature of this algebra: the presence of terms including
 ${\cal G}_I {\cal H}_{\mu}$.
This leads to the possibility of two different forms of the BRST-charge of the
theory.

Introducing a Grassmann-odd ghost-variable to each constraint of a theory
one can always find a nilpotent quantity called the BRST-charge \cite{teit}.
 A general form of the BRST-charge was given for example in \cite{Marnelius}.

\begin{equation}
Q = {\Psi}_a \eta^a + \sum_{i=1}^{N} U^{a_1...a_i} {\cal P}_{a_1}...
 {\cal P}_{a_i}
\end{equation}
where N is the rank of the theory, $\Psi$-s are the constraints and
 ${\cal P}_{a_i}$
are the ghost-momenta canonical to the $\eta^a$ ghost-variables. Knowing the
constraint-algebra the coeficients $ U^{a_1...a_i}$ can be computed.
The first ones are especially simple:

If $ \{ {\Psi}_a,{\Psi}_b \} = C_{ab}^{~~c} {\Psi}_c $ then
$ U^c = -\frac12  C_{ab}^c \eta^a \eta^b $ and from here:
\begin{equation}
Q = {\Psi}_a \eta^a -\frac12  C_{ab}^c \eta^a \eta^b {\cal P}_c + ...
\end{equation}
If we now look at the algebra of the constraints ${\cal H}_{\mu}$
(equations~\ref{alg1}  and \ref{alg2}) we notice
that the coeficients $U^c$ can be written in two different ways.

The term
 \[ {\cal H}_{\mu} {\cal G}_I [{\frac12}{\epsilon}_{ab}(M^{-1})^{\mu I}
 P^a N^b] \]
can be understood either as:

CHOICE 1: $ C_{ab}^I {\cal G}_I$, that is the Poisson bracket between the two
constraints equals to Gauss' law multiplied by a coefficient that contains
a constraint:
 $ C_{ab}^I = {\frac12}{{\epsilon}_{ab}} {\cal H}_{\mu}  (M^{-1})^{\mu I}$
or

CHOICE 2:  $ C_{ab}^{'\mu} {\cal H}_{\mu}$ which means  that the P.B.
 is a constraint ${\cal H}_{\mu}$ times a coefficient containing Gauss' law.
 This coefficient is:
  $ C_{ab}^{'\mu}= {\frac12}{\epsilon}_{ab} {\cal G}_I (M^{-1})^{\mu I}$

This means that there are two possible forms of the BRST-charge, depending
on wich way one wants to understand the algebra.

   The algebra of the $2+1$ dimensional gravity is of course the simplest
if one uses the constraints suggested by Witten \cite{Witten}.
 This gives the simplest form to the BRST-charge too:

\begin{equation}
Q^{W}= {\Psi}^I \tilde{\eta}_{I} +{\cal G}_I \tilde{\rho}^{I} - f^{I~J}_{~K}
\tilde{\eta}_{I}
\tilde{\rho}^{K} \tilde{\cal P}_{J} - \frac12  f^I_{JK}\tilde{\rho}^{J}
\tilde{\rho}^{K}
 \tilde{\cal P}_{I}
\end{equation}
where
  $\tilde{A}^J_b$ and $\tilde{E}^a_I $ are the phase-space variables,
 $\tilde{\cal P}_I $ -the momentum canonical to the ghost $\tilde{\rho}^I$ ,
  $\tilde{\Pi}^{I} $ -the momentum canonical to the ghost $\tilde{\eta}_{I} $.

This is thus a rank 1 form of the BRST-charge.

There is a theorem due to Batalin and Fradkin \cite{fradkin}
 that says: if a BRST-charge can
be expressed in two different forms there exists a canonical transformation
in the extended phase space
relating the variables used in the two expressions. This means that one should
 be able
to find a canonical transformation that relates  $\tilde{A}^{J}_b $,
  $\tilde{E}^{a}_I $,
 $\tilde{\rho}^{I} $, $\tilde{\cal P}_{I} $, $\tilde{\eta}_I $ and
 $\tilde{\Pi}^I $
to the variables $ A^I_a$,$ E^a_I$, $\rho^I$,$ {\cal P}_{I} $ , ${\eta}_{I}$
and ${\Pi}^{I}$ used in the expression of the two forms of $ Q^{(A)}$.
 The first natural
 thing is to try to find a set of canonical transformations starting with:
 \begin{equation}
 \tilde{\eta}_I = M_{\mu I} {\eta}^{\mu}
 \end{equation}
This gives of course:
 \begin{equation}
 \tilde{\Pi}^I = (M^{-1})^{\mu I} {\Pi}_{\mu}
\end{equation}
The other relations between these variables are:
\begin{equation}
 \tilde{A}^I_a = {A}^I_a - f^{I~K}_{~J}(M^{-1})^{\mu J} M_{a K} {\Pi}_{\mu}
\eta -
\frac{{\epsilon}_{ab}}2(M^{-1})^{\mu I} {\Pi}_{\mu} {\eta}^b
\end{equation}
while the other three variables remain unchanged: $\tilde{E}^a_I =  E^a_I$,
$ \tilde{\rho}^K = {\rho}^K$ and $\tilde{\cal P}_J =  {\cal P}_J$

One should stress again that $\mu$ is not a vector index in this case.
The BRST-charge computed by substituting the variables with tilde in
 $ Q^{(W)} $ by
$ E^a_I $,$ {A}^I_a$, etc. is:

\[ Q^{(A2)} = {\cal H} \eta + {\cal H}_a {\eta}^a +{\cal G}_I [{\rho}_I -
 \frac{{\epsilon}_{ab}}2  (M^{-1})^{b I} \Pi \eta {\eta}^a \]
\[ +  \frac{{\epsilon}_{ab}}2 (M^{-1})^{0 I} q^{db} {\Pi}_d \eta {\eta}^a
+ \frac{{\epsilon}_{ab}}4 (M^{-1})^{\mu I} {\Pi}_{\mu} {\eta}^a  {\eta}^b] \]
\[+\Pi ({\partial}_c {\eta}^c \eta -  {\eta}^c {\partial}_c \eta )
- q_{cd} {\Pi}_d {\partial}_c \eta \eta \]
 \begin{equation}
+{\partial}_c ( {\Pi}_d {\eta}^c ) {\eta}^d
- \frac12 f_I^{~JK} {\rho}_J {\rho}_K {\cal P}^I
\end{equation}
This corresponds to the "second choice" of the coefficients on page 3 and it
is of rank 1. One can of course compute the form of the BRST-charge
 corresponding
to the first choice. This is done in the simplest way as in \cite{ash2}. We
get:
\[ Q^{(A1)} = {\cal G}_I{\rho}^I +
{\cal H}_{\mu} [{\eta}^{\mu} + (M^{-1})^{\mu I} \frac{{\epsilon}_{ab}}4
 {\cal P}_I {\eta}^a{\eta}^b \]
\[ -(M^{-1})^{\mu I} \frac{{\epsilon}_{ab}}2 f_{I}^{~JK} E^b_J
 {\cal P}_K \eta {\eta}^a \]
\[ + (M^{-1})^{\mu I}\frac{{\epsilon}_{ab}}4  f_{I}^{JK} {\cal P}_J {\cal P}_K
\eta {\eta}^a  {\eta}^b] \]

\[ - {\frac12} f_{IJ}^{~~K} {\rho}^I {\rho}^J {\cal P}_K
-{\Pi}_d {\eta}^c {\partial}_c  {\eta}^d
-\Pi {\eta}^c {\partial}_c  {\eta}
+\Pi {\partial}_c  {\eta}^c \eta \]
\begin{equation}
-q^{cd} {\Pi}_d \eta  {\partial}_c  {\eta}
-2E^I_b {\cal P}_I {\Pi}_b \eta {\partial}_c  {\eta} {\eta}^c
\end{equation}

As we see this corresponds to rank 2. There exists of course a transformation
between $Q^{(A1)}$ and $Q^{(A2)}$ and another one between $Q^{(A1)}$  and
 $Q^{(W)}$
but they are very complicated.

 We obtained three different forms of the BRST-charge of the same
 theory, two of them being of rank 1 ( $Q^{W}$ and $Q^{A2}$) and one of them
rank 2 ($Q^{A1}$). This is an interesting example of the fact that a theory can
have BRST-charges of different rank depending on how one formulates the
constraints so one can change the rank of a theory by reformulating the
constraints \cite{fradkin}, \cite{teit}.

 \paragraph{ACKNOWLEDGEMENTS}

 I am grateful to Ingemar Bengtsson for suggesting me these questions.
I want also to thank Peter Peld\'an for interesting remarks.

\end{document}